\documentclass[aps,twocolumn,groupedaddress]{revtex4}
\usepackage{graphicx}
\usepackage{amssymb}
\begin{document}
\title{Ultracold atoms and Bose-Einstein condensation for quantum metrology}
\author{H{\'e}l{\`e}ne Perrin}
\affiliation{Laboratoire de physique des lasers, CNRS-Universit{\'e} Paris-Nord, 99 avenue J.-B. Cl{\'e}ment, 93430 Villetaneuse, France}
\begin{abstract}
This paper is a short introduction to cold atom physics and Bose-Einstein condensation. Light forces on atoms are presented, together with laser cooling, and a few atom traps: the magneto-optical trap, dipole traps and magnetic traps. A brief description of Bose-Einstein condensation is given together with some important links with condensed matter physics. The reader is referred to comprehensive reviews and to other lecture notes for further details on atom cooling, trapping and Bose-Einstein condensation.
\end{abstract}
\maketitle
\section{Introduction}
\label{sec:intro}
The field of ultra cold atoms and Bose-Einstein condensation has developed dramatically since the first proposal of laser cooling in the 70's\cite{Hansch1975}. High precision spectroscopy is a first natural application of atom cooling, since Doppler-free lines become observable. Freezing the atomic motion is an essential tool in modern frequency-time metrology, in atomic fountains as well as in optical clocks~\cite{QMBize,QMLemonde,QMMargolis}. Reaching temperatures of one microkelvin or less gives also access to experimental quantum physics with atoms. Regarding single particle properties, ultracold atoms make interferometry experiments with atoms easier, taking advantage of their large de Broglie wavelength $\lambda = \displaystyle {h}/{(Mv)}$, where $M$ is the atomic mass and $v$ the velocity. Atom interferometry is now widely using cold atoms, in particular for applications to quantum metrology~\cite{QMGuellati}. When the density of the atoms gets larger, depending on their bosonic or fermionic nature, ultracold atoms lead to the observation of Bose-Einstein condensation in atomic vapours \cite{Anderson1995,Cornell2002,Ketterle2002} or degenerate Fermi seas \cite{DeMarco1999}. This happens when the atomic density $n$ and the thermal de Broglie wavelength $\lambda_{\rm dB} = h/\sqrt{2\pi M k_B T}$ reach a critical value:
\begin{equation}
		n \lambda_{\rm dB}^3 \gtrsim 1 .
\end{equation}
$T$ denotes the gas temperature. Degenerate quantum gases, explored in dilute gases since 1995, have given rise to a tremendous series of experimental results, witnessing the rich physics of this rapidly growing field, including coherence properties, superfluidity, quantum phase transitions, and correlations. They are excellent candidates for quantum computation. They also bring new insights in many areas of condensed matter physics: Bloch oscillations, superfluid-insulator transitions, Cooper pairing and Josephson oscillations (giving rise to wide applications in quantum
electrical metrology~\cite{QMJeanneret,QMKemppinen,QMGallop}), search for Anderson localization... (which was recently observed in momentum space in Cs clouds~\cite{Chabe2007} and in Bose-Einstein condensate of Rb~\cite{Billy2008}). The two communities are benefiting from cross-fertilisation of complementary ideas.

This paper will not give a comprehensive course on ultracold atom physics, as many detailed lecture notes are already available, in particular from former summer schools dedicated to the subject~\cite{LesHouches90,Varenna98,Dalfovo1999,LesHouches99,Pethick2002,Pitaevskii2003}. Instead, we will give a brief overview of these Nobel prize winning topics and the reader is encouraged to consult the references given at the end of the notes.
\section{Light forces on atoms}
\label{sec:lightforces}
For a detailed review of light forces on atoms, see the Les Houches lectures of Cohen-Tannoudji \cite{LesHouches90}, his lectures at Coll\`ege de France~\cite{CohenCollege} or the book of Metcalf and van der Straten~\cite{Metcalf}. One can also consult the Nobel lectures of Cohen-Tannoudji~\cite{CohenNobel}, Chu~\cite{ChuNobel} and Phillips~\cite{PhillipsNobel}.
\subsection{Orders of magnitude}
\label{sec:numbers}
In this section, we restrict ourselves to the two level approximation, with an energy difference $\hbar \omega_0$ between a ground state $| g \rangle$ and an excited state $| e \rangle$. Light forces occur due to repeated momentum changes of the atom at each absorption or emission of a photon. For each near resonant photon of frequency $\omega$ absorbed or emitted, the atomic velocity is changed due to momentum conservation by the \textit{recoil velocity}
\begin{equation}
v_{\rm rec} = \frac{\hbar k}{M}
\label{eq:vrec}
\end{equation}
where $k=2\pi/\lambda=\omega/c$ and $ \lambda$ is the wavelength of the incoming light. The order of magnitude of the recoil velocity is 3 to 30\,mm$\cdot$s$^{-1}$ for alkali atoms. Photon scattering occurs at a typical rate $\Gamma$, the inverse lifetime of the excited state, of the order of $10^7$ to $10^8$\,s$^{-1}$. As a result, the typical acceleration undergone by alkali atoms in near resonant light is $10^4$ to $10^5$ times the earth acceleration $g$, which explains the great success of laser manipulation of atomic external degrees of freedom.

\subsection{Atom - light interaction}
\label{sec:atomlight}
Let us consider an atom in the field of monochromatic laser light. Three interacting systems have to be considered: the laser modes, the two level atom of hamiltonian $\hat{H}_{\rm at}$ and the vacuum modes of the field, hamiltonian $\hat{H}_{\rm vac}$. The atom interacts with light through its induced electric dipole moment $\mathbf{\hat{D}}$, coupled with the light field operator through $\hat{V}=-\mathbf{\hat{D}}\cdot\mathbf{\hat{E}}$. The interaction $\hat{V}_{\rm vac}$ of the atom with the continuum of the vacuum modes is described by a finite lifetime $\Gamma^{-1}$ of the excited state, resulting from the application of the Fermi golden rule. These spontaneous emission processes are essential for allowing another absorption. However, they do not contribute to the average force felt by the atoms, due to the randomness of the direction of emission. By contrast, their fluctuations are important for evaluating the limit temperature in laser cooling.

The  classical field of a laser
\begin{equation}
\mathbf{E}(\mathbf{r},t) = \mathcal{E}(\mathbf{r})/2 \times \left( \mathbf{\epsilon}(\mathbf{r}) e^{-i \omega t - i \phi(\mathbf{r})} + \mbox{c.c.}\right)
\end{equation}
is coupled to the atomic electric dipole moment $\mathbf{d}=\langle e | \hat{\mathbf{D}} | g \rangle$ through the Rabi frequency $\Omega$, where
\begin{equation}
\Omega(\mathbf{r}) = - \left(\mathbf{d}\cdot\mathbf{\epsilon}(\mathbf{r})\right) \mathcal{E}(\mathbf{r}) .
\end{equation}
The laser polarisation $\mathbf{\epsilon}(\mathbf{r})$, amplitude $\mathcal{E}(\mathbf{r})$ and phase $\phi(\mathbf{r})$ have been introduced here. The Rabi frequency is related to the laser intensity by defining a saturation intensity $I_s$, such that
\begin{equation}
\frac{\Omega^2}{\Gamma^2} = \frac{I}{2I_s} .
\end{equation}
For alkali atoms on the dipolar cycling transition, the saturation intensity is of the order of a few mW/cm$^3$. This gives an order of magnitude of the intensity necessary for pushing atoms at resonance. This value is reasonably low, such that low power laser diodes may be used for realising a magneto-optical trap, discussed in section~\ref{sec:MOT}.

Near resonance, that is if the detuning $\delta = \omega - \omega_0$ is small, $|\delta| \ll \omega_0$, the atom-laser coupling term may be written in the rotating wave approximation (RWA)
\begin{equation}
\hat{V}_{\rm laser} = - \hat{\mathbf{D}}\cdot\mathbf{E}(\hat{\mathbf{r}},t)\simeq \frac{\Omega(\hat{\mathbf{r}})}{2}\left( | e \rangle \langle g | e^{-i \omega t - i \phi(\hat{\mathbf{r}})} + \mbox{h.c.} \right)
\end{equation}
The mean force $ \langle \hat{\mathbf{F}}\rangle$ acting on an atom for a given position $\mathbf{r}$ and velocity $\mathbf{v}$ is obtained in the Heisenberg picture by averaging over the internal variables, which evolve much faster than the external ones:
\begin{equation}
\mathbf{F} = \langle \hat{\mathbf{F}}\rangle = \langle \frac{d\hat{\mathbf{P}}}{dt} \rangle = \frac{1}{i\hbar} \langle \left[ \hat{\mathbf{P}},\hat{H} \right] \rangle = - \langle \nabla \hat{V}_{\rm laser} \rangle 
\end{equation}
where $\hat{\mathbf{P}}$ is the atomic momentum and $\hat{H} = \hat{H}_{\rm at} + \hat{H}_{\rm vac} + \hat{V}_{\rm vac} + \hat{V}_{\rm laser}$ is the hamiltonian describing both the atom and the field. The mean contribution of the coupling to vacuum $\hat{V}_{\rm vac}$ is zero, as already mentioned. The fluctuation of the mean force is responsible for the momentum diffusion. The gradient of atom-laser coupling may come from an intensity gradient or a phase gradient of the laser field.

\subsection{The light forces}
\label{sec:forces}

\subsubsection{Radiation pressure}
The \textit{radiation pressure} arises from a phase gradient. The typical example is the plane wave, for which $\phi(\mathbf{r}) = - \mathbf{k\cdot r}$. Then,
\begin{eqnarray}
\mathbf{F}_{\rm pr} &=& \hbar \mathbf{k} \frac{\Gamma}{2} \frac{s}{1+s} \mbox{ , \quad where }\\
 s &=& \frac{\Omega^2/2}{\delta^2 + \Gamma^2/4} = \frac{I/I_s}{1 + 4\delta^2/\Gamma^2} . \nonumber
\end{eqnarray}
$s$ is the \textit{saturation parameter} and may depend on the position $\mathbf{r}$ if $I$ does.

The interpretation of the radiation pressure is straightforward. Absorption -- spontaneous emission cycles occur at a rate $\gamma_{\rm fluo}$, where $\gamma_{\rm fluo} = \frac{\Gamma}{2} \frac{s}{1+s}$. At each cycle, the atomic momentum changes on average by $\hbar k = M v_{\rm rec}$ in the direction of the plane wave, as spontaneous emissions in opposite directions are equally probable. The resulting force is then $\mathbf{F}_{\rm pr} = \hbar \mathbf{k}  \gamma_{\rm fluo} $. 

The maximum value of the radiation pressure is $\mathbf{F}_{\rm pr} = \hbar \mathbf{k}  \frac{\Gamma}{2}$. For sodium atoms, this corresponds to an acceleration $a \sim 10^5 \, g$. This force is thus able to stop atoms from a thermal beam initially  at $v=100$\,m/s over a distance of 1\,cm! 

\subsubsection{Example of application: the Zeeman slower}
An important application of this large force is the Zeeman slower, first demonstrated by W.D.~Phillips, H.~Metcalf and their colleagues~\cite{Prodan1985}. The basic idea is to use a laser propagating against an atomic beam to slow it down to almost zero velocity, over a length $d$ of the order of one meter. The radiation pressure force is well adapted to this purpose. However, the scattering rate $ \gamma_{\rm fluo}$ depends on the detuning between the laser frequency and the atomic transition. As the atomic velocity changes, the atomic transition is shifted by the first order Doppler effect: $\delta = \omega - \omega_0 - \mathbf{k}\cdot \mathbf {v}$. For maintaining the resonance condition $|\delta| < \Gamma$,  Phillips and Metcalf proposed to compensate this shift by an opposite Zeeman shift.

In the presence of a magnetic field, the magnetic sublevels are shifted by an amount proportional to the magnetic field $B$, the magnetic dipole moment $\mu$ and the spin projection $m$. This implies a description of the atomic transition beyond the two-level model. The simplest situation is a $J=0 \longrightarrow J'=1$ transition. The Zeeman effect shifts the atomic transition to $|J'=1,m'\rangle$ by $m' \gamma B$ where $\gamma = \mu/\hbar$ is the gyromagnetic factor. With a $\sigma^+$ polarised laser, only the transition to $m'=+1$ is allowed and the magnetic field profile is adjusted such that the light remains resonant as the atomic velocity decreases: $\omega_0(z) = \omega_0 + \gamma B(z) = \omega + k v(z)$. Typically, $B$ should decrease like $\sqrt{1-z/d}$.

\subsubsection{Dipole force}
\label{sec:dipoleforce}
The \textit{dipole force} results from an intensity gradient.
\begin{equation}
\mathbf{F}_{\rm dip} = -\frac{\hbar\delta}{2} \frac{\nabla s(\mathbf{r})}{1+s(\mathbf{r})}
\end{equation}
It derives from the dipole potential $U_{\rm dip}(\mathbf{r}) = \displaystyle \frac{\hbar\delta}{2} \ln \left( 1+s(\mathbf{r}) \right)$.

The dipole force comes from photon redistribution inside the laser beam: it is thus based on stimulated emission, in contrast to the radiation pressure based on spontaneous emission.
The dipole force is zero on resonance, but becomes important when the laser is detuned. For detunings larger than the natural width $|\delta| \gg \Gamma$ and in the low saturation limit $s \ll 1$, the dipole potential simplifies to
\begin{equation}
U_{\rm dip} = \displaystyle \frac{\hbar\Omega^2}{4\delta}.
\label{eq:Udip}
\end{equation}

Depending on the sign of the detuning $\delta$, two situations occur:
\begin{itemize}
\item[$\delta<0$] \textit{red detuning}: The potential is minimum where the light intensity is maximum: light acts as an attractive potential for the atoms.

Example: the focus point of a red detuned Gaussian laser beam is a 3D trapping potential for neutral atoms, as Chu  and his colleagues demonstrated in 1986~\cite{Chu1986}.
\item[$\delta>0$] \textit{blue detuning}: The potential is maximum where the light intensity is maximum: light acts as a repulsive potential for the atoms.

Example: a blue detuned evanescent wave at the surface of a dielectric acts as an atomic mirror, atoms being repelled from the high intensity region close to the surface\cite{Cook1982,Balykin1987}.
\end{itemize}

Finally, the spontaneous emission rate $\gamma_{\rm fluo}$ scales as $\Gamma \frac{\Omega^2}{\delta^2} \propto \frac{\Gamma}{\delta} U_{\rm dip}$. The dipole force is thus conservative for large detunings, with a vanishing spontaneous scattering rate, which makes this force well suited for implementing conservative traps. For a detailed review of dipole traps, one may consult the reference~\cite{Grimm2000}. An important application of dipole traps is the optical lattice, where light standing waves --- either blue or red detuned --- create a periodic potential for the atoms in one, two or three dimensions. Optical lattices were first studied with near resonant light and thermal atoms~\cite{Westbrook1990,Grynberg2001}. Conservative optical lattices are now widely used with Bose-Einstein condensates to mimic solid state physics problems, with a control over filling fraction, potential depth, tunnelling to atom interaction ratio and effective mass~\cite{Bloch2005}. Alternatively, the strong confinement of optical lattices is used in recent atomic optical clocks~\cite{QMBize}.

\section{Laser cooling}
\label{sec:lasercooling}
Laser cooling relies on the light forces exerted by a laser onto atoms, as in the case of the Zeeman slower. The simplest mechanism is Doppler cooling. Other cooling mechanisms allow the observation of lower temperatures but their detailed discussion is beyond the scope of this paper.

\subsection{Doppler cooling}
\begin{figure}
\begin{center}
\includegraphics[width=60mm]{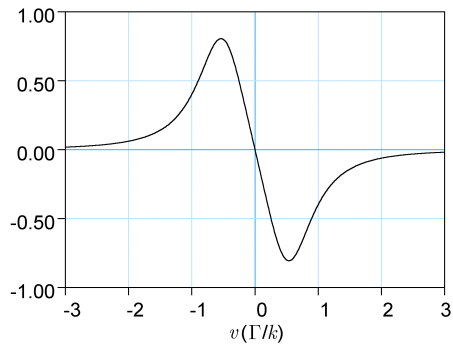}
\caption{Doppler force in units of $\hbar k \Gamma s_0$, as a function of atomic velocity in units of $\Gamma/k$. Below $\Gamma/2k$, the force is almost linear, it is a friction force.}
\label{fig:Dopplerforce}       
\end{center}
\end{figure}

Doppler cooling was suggested by H\"ansch and Schawlow in 1975~\cite{Hansch1975}. 
Let us consider a moving atom, with velocity $\mathbf{v}$, in the field of two counter-propagating red detuned laser beams, with detuning $\delta = \omega - \omega_0$, saturation intensity $s_0$ each and wave vector $\mathbf{k}$ and $-\mathbf{k}$. Due to the Doppler shift, the atom sees the two lasers with a different frequency $\omega \pm \mathbf{k}\cdot\mathbf{v}$. In the low intensity limit $s \ll 1$, the radiation pressure forces of the two beams add, see Fig.\,\ref{fig:Dopplerforce}:
\begin{eqnarray}
\mathbf{F} &=& \hbar\mathbf{k} \frac{\Gamma}{2} \left( s_+(\mathbf{v}) - s_-(\mathbf{v}) \right) \\
s_{\pm}(\mathbf{v}) &=& \frac{I/I_s}{1+4(\delta \mp \mathbf{k\cdot v})^2/\Gamma^2} . \nonumber
\end{eqnarray}

At low velocity $v \ll \Gamma/k$, one gets a \textit{friction force} $\mathbf{F} = -\alpha \mathbf{v}$ with a friction coefficient
\begin{equation}
\alpha = \hbar k^2 \, s_0 \frac{-2\delta\Gamma}{\delta^2 + \Gamma^2/4} .
\end{equation}

The detuning must be negative to have $\alpha > 0$. The friction coefficient is maximum $\alpha_{\rm max} = 2 \hbar k^2 \, s_0$ for $\delta=-\Gamma/2$.The velocity capture range of Doppler friction is of order $v_{\rm capt} = \Gamma/k$, a few to a few tens of meters per seconds for alkali atoms, such that atoms can be cooled directly from a thermal distribution at room temperature. The velocity damping time is of order $\hbar/E_{\rm rec}$ where $E_{\rm rec} = Mv_{\rm rec}^2/2$ is the recoil energy, that is a few 100\,$\mu$s for Rb at $s_0=0.1$. This extremely powerful scheme also works in 3D with three pairs of counter-propagating beams. Kinetic temperatures of 1~mK or below are reached within 1~ms typically. Such a cooling scheme is called an \textit{optical molasses}, and was demonstrated for the first time in 1985 by Chu and his colleagues~\cite{Chu1985}.

The limit temperature that can be achieved with Doppler cooling is given by the competition betwen the friction force and a random walk in momentum space induced by the randomness of spontaneous emission. As explained above, the mean force due to the coupling to vacuum modes is zero, however the fluctuations of this force limit the final temperature of Doppler cooling. The diffusion coefficient in momentum is $D_p = \hbar^2k^2\Gamma s_0$ \cite{Letokhov1977,LesHouches90} and leads to the temperature $k_B T = D_p/\alpha$. The temperature is minimum for the detuning $\delta=-\Gamma/2$ where $\alpha = \alpha_{\rm max}$. This limit temperature is called the \textit{Doppler temperature} $T_D$:
\begin{equation}
k_B T_{D} =  \frac{\hbar\Gamma}{2}.
\end{equation}
This temperature corresponds to $240\,\mu$K for sodium, $125\,\mu$K for caesium.

\subsection{Sub-Doppler cooling}
Soon after the first experimental demonstration of an optical molasses, precise measurements of the atomic cloud temperature revealed that the observed temperature $40\,\mu$K was lower than the $240\,\mu$K predicted for sodium by the Doppler cooling theory~\cite{Lett1988}. Moreover, the temperature was found to depend on laser intensity and detuning as $I/|\delta|$, again in contradiction with theory which predicted no dependence with intensity and a minimum temperature at $\delta=-\Gamma/2$.

\begin{figure}
\begin{center}
\includegraphics[width=0.8\columnwidth]{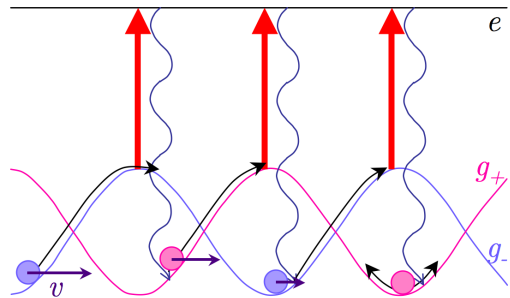}
\caption{Sisyphus cooling. Depending on their internal state $g_+$ or $g_-$, the atoms see two periodic potentials of opposite phase. Due to optical pumping through the excited state $e$, atoms are more often climbing hills than accelerating down.}
\label{fig:Sisyphus}       
\end{center}
\end{figure}

The explanation to these exciting results came one year after~\cite{Dalibard1989}: the light polarisation together with the sublevel structure of the ground state and the excited state play an important role in the cooling mechanism. Due to interferences between the different cooling beams, the polarisation varies in space on the scale of $\lambda/2$, creating a polarisation lattice. The light shift of the Zeeman sublevels, resulting from the dipole potential, is negative and depends on polarisation. The atoms then see a periodic conservative dipole potential depending on their internal state. On the other hand, optical pumping, a dissipative effect, favours the population into the most shifted states, which are the lowest in energy. As they move inside the molasses, atoms thus lose energy through the pumping processes and accumulate at the bottom of the lattice wells. This cooling mechanism is called Sisyphus cooling --- or polarisation gradient cooling --- as the atoms always climb on hills, losing kinetic energy, and are put back at the bottom of the hill by a dissipative pumping process~\cite{LesHouches90}. A sketch of this mechanism is given in Fig.\,\ref{fig:Sisyphus}.

The process stops when the atoms have not enough energy to climb the next hill. As the lattice depth is given by the dipole potential, proportional to $I/\delta$, see Eq.(\ref{eq:Udip}), we recover the fact that the final temperature scales as $I/|\delta|$. The limit temperature is found to be a few times the \textit{recoil temperature} $T_{\rm rec} = 2 E_{\rm rec}/k_B$. This is due to the cooling process itself, based on photon scattering, which limits the final velocity to a few times the elementary step in velocity diffusion, the recoil velocity $v_{\rm rec}$, see Eq.(\ref{eq:vrec}). The recoil temperature is usually much less than the Doppler temperature, for example $2.4\,\mu$K for sodium and 200\,nK for caesium.

\subsection{Sub-recoil cooling}
To go beyond the recoil temperature with laser cooling, the spontaneous processes must be made velocity dependent, in order to protect the coldest atoms from light scattering. This may be achieved either using an interference in the coupling to light~\cite{Aspect1988} or by using velocity selective excitations as in Raman cooling~\cite{Kasevich1992}.

On the other hand, if the atoms are trapped in a very confining conservative potential, such that the level spacing between eigenstates $n$ is larger than the recoil energy, sideband cooling is a very efficient cooling technique~\cite{Wineland1975}. A transition between $|g,n\rangle$ and $|e,n-1\rangle$ is driven by a laser tuned to the red sideband, whereas the spontaneous decay keeps $n$ constant. After an absorption-emission cycle, the atomic energy has decreased by the level spacing. When this procedure is repeated, atoms accumulate into the trap ground state~\cite{Diedrich1989,QMMargolis}.

Finally, a cooling technique alternative to laser cooling is now widely used to produce Bose-Einstein condensates: evaporative cooling, implemented in conservative traps, see section~\ref{sec:evap}.

\section{Traps for neutral atoms}
\label{sec:traps}
Together with laser cooling, trapping techniques were developed to improve the observation time. Ions can be trapped very efficiently with electric fields due to their electric charge, and sideband cooling is implemented in ion traps. These trapping and cooling schemes are reviewed in the paper of Helen Margolis~\cite{QMMargolis}, and I will concentrate here on traps for neutral atoms. These traps are either dissipative, like the magneto-optical trap, or conservative, like dipole traps and magnetic traps.

\subsection{The magneto-optical trap}
\label{sec:MOT}
The realisation of the first magneto-optical trap (MOT) was a very important step toward the achievement of Bose-Einstein condensation. It allows, in a simple and quick step --- at least for alkali atoms --- to increase the phase space density from about $10^{-19}$ in a vacuum chamber at 300\,K to $10^{-7}$ in a fraction of a second. Demonstrated for the first time in 1987 by Raab \textit{et al.}~\cite{Raab1987} from a suggestion of Dalibard, it became quickly very popular and is nowadays even used in some teaching labs for undergraduate students.

\begin{figure}
\begin{center}
\includegraphics[width=0.6\columnwidth]{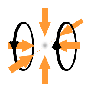}

\begin{tabular}{ll}
$J=0$ ~ \includegraphics[width=0.6\columnwidth]{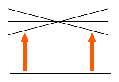}&$m=0$\\[-35mm]
&$m'=1$\\[3mm]
$J'=1$&$m'=0$\\[3mm]
&$m'=-1$\\[3mm]
~ \hspace{10mm} $\sigma^+$ \hspace{25mm} $\sigma^-$ &
\\[5mm]
&
\end{tabular}
\caption{Principle of the magneto-optical trap. Top: the MOT consists of three pairs of counter-propagating beams with opposite $\sigma^+/\sigma^-$ polarisation plus a pair of coils with opposite current for the magnetic field gradient. Bottom: polarisation-dependent coupling to the excited states, shifted by the magnetic field gradient (1D scheme).}
\label{fig:MOT}       
\end{center}
\end{figure}

The idea is to combine the optical molasses with a magnetic field gradient such that the detuning --- and thus the photon scattering rate from each laser --- depends both on atomic velocity \textit{and} position. We consider the simplest case of a $J=0 \longrightarrow J'=1$ transition. In the presence of a magnetic field $B(z)=b'z$, the excited magnetic sublevels $|J'=1,m'\rangle$ are shifted by $m' \hbar \gamma b'z$, see Fig.~\ref{fig:MOT}. The laser polarisation of the beam propagating towards $+z$ is chosen to be $\sigma^+$. The light field is thus exciting the $|J=0,m=0\rangle \longrightarrow |J'=1,m'=1\rangle$ transition, with resonant frequency $\omega_0 + \gamma b'z$. Taking into account the atomic velocity $\mathbf{v}$, the effective detuning to resonance is $\delta' = \omega - \mathbf{k\cdot v} - (\omega_0 + \gamma b'z) = \delta - \mathbf{k\cdot v} - \gamma b'z$. Symmetrically, the counter-propagating beam has a $\sigma^-$ polarisation exciting the $|J=0,m=0\rangle \longrightarrow |J'=1,m'=-1\rangle$ transition. In the low saturation limit, the radiation pressure forces of the two counter-propagating beams add and the total force reads:
\begin{eqnarray}
\mathbf{F} &=& \hbar\mathbf{k} \frac{\Gamma}{2} \left( s_+(\mathbf{v},z) - s_-(\mathbf{v},z) \right) \\
s_{\pm}(\mathbf{v},z) &=& \frac{I/I_s}{1+4(\delta \mp \mathbf{k\cdot v} \mp \gamma b'z)^2/\Gamma^2} \: . \nonumber
\label{eq:MOT}
\end{eqnarray}
Atoms are attracted towards the position $z=0$. A generalisation to 3D is straightforward using two coils in an anti-Helmholtz configuration for the magnetic field\footnote{The magnetic gradient is twice as large and with opposite sign along the coil axis, such that the laser polarisation must be reversed and the atomic cloud is elliptical if the laser intensity of all the six beams is the same.}. At low velocity and close to the magnetic field zero, Eq.(\ref{eq:MOT}) can be linearised and the friction force is now completed by a restoring force $-\kappa \mathbf{r}$ where
\begin{equation}
\kappa = \hbar k \gamma b' \, s_0 \frac{-2\delta\Gamma}{\delta^2 + \Gamma^2/4} \: .
\end{equation}

At low atomic density, the equation of motion is the one of a damped oscillator. The cloud size is proportional to the square root of the temperature and the density has a Gaussian profile. At larger density however, a Coulomb-like repulsive force between atoms appears, due to reabsorbed scattered photons: a photon emitted by an atom $A$ is reabsorbed by an atom $B$, resulting in a relative momentum change of $2 \hbar k$, that can be modelled as a repulsive force. The interplay between the restoring force and the repulsive force results in a uniform atomic density
\begin{equation}
n_0 = \frac{16\pi}{3} \frac{\mu b' |\delta|}{\hbar \lambda^2 \Gamma^2}
\end{equation}
inside a sphere of radius $R_{\rm MOT} \propto (N/n_0)^{1/3} $. This effect limits the density in a magneto-optical trap to a few $10^{10}$\,cm$^{-3}$ typically. To increase the density further, atoms should be pumped into a dark state that does not interact with light.

In the trap, the temperature commonly reaches $100 \, T_{\rm rec}$, with a phase space density $10^{-7}$ to $10^{-6}$. It is rather easy to implement, at least for alkali atoms, as atoms can be loaded directly from an atomic vapour. All these features make the magneto-optical trap a very popular tool in modern atomic physics.

\subsection{Conservative traps}
Although the MOT is very powerful, the phase space density in a MOT is still orders of magnitude too low for reaching quantum degeneracy. This is related to the photon scattering process, which is always present and limits both the temperature and the density. Even if better results can be achieved in dedicated experiments\,\cite{Kerman2000,Han2000}, another approach consists in loading atoms in a conservative trap and performing evaporative cooling. Up to now, this is the only method that leads to Bose-Einstein condensation.

\begin{figure}
\begin{center}
\includegraphics[width=0.8\columnwidth]{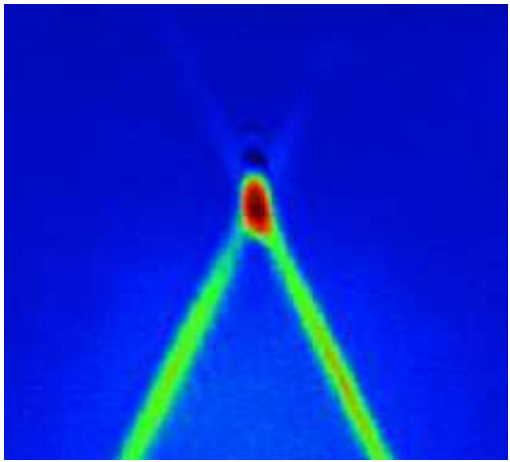}
\caption{A conservative trap is realised by crossing two, far off resonant, red detuned laser beams. The trap is loaded from a magneto-optical trap. Caesium atoms that were not initially at the crossing fall due to gravity, preferentially along the axes of the laser beams.}
\label{fig:dipoletrap}       
\end{center}
\end{figure}

Dipole traps were presented in section\,\ref{sec:dipoleforce}. The simplest case is the focal point of a red detuned laser~\cite{Chu1986}, which attracts the atoms towards the maximum of light intensity. To improve the confinement, two crossed beams may be used, see Fig.\,\ref{fig:dipoletrap}. As the photon scattering rate scales as $I/\delta^2$ whereas the light shift scales as $I/\delta$, with $I$ the light intensity and $\delta$ the detuning, the use of high intensity together with large detunings allows the realisation of atom traps where spontaneous photon scattering is suppressed, with sufficient depth to be loaded from a magneto-optical trap. Bose-Einstein condensation of rubidium was observed in an all-optical trap in 2001, by reducing the intensity of a CO$_2$ trapping laser~\cite{Barrett2001}.

Another popular conservative trap widely used for reaching quantum degeneracy is the magnetic trap. It makes use of the magnetic dipole moment $\mu$ of the atom. An inhomogeneous static magnetic field $B(\mathbf{r})$ is realised with current flowing into several coils, such that a minimum of magnetic field is produced. If the atomic spin is anti-parallel to the local magnetic field everywhere, the atoms see a potential $ V(\mathbf{r}) = \mu |B(\mathbf{r})| $ and are trapped to the vicinity of the field minimum. Note that if the magnetic field traps the atoms in magnetic spin states $|F,m_F\rangle$ (the low-field seekers), the atoms in state $|F,-m_F\rangle$ (the high-field seekers) see an ``anti-trap'' and are expelled from the low field region.

The simplest magnetic trap is the quadrupole trap, resulting from the quadrupolar field produced by two coils of same axis with opposite currents. The drawback of such a trap is that the field vanishes at the centre, leading to spin flips -- or Majorana losses. To produce a non zero field minimum, a typical solution is given by the Ioffe-Pritchard trap~\cite{Pritchard1983}. Four parallel bars placed on a square with alternate current create a transverse quadrupole field, and two coils at the extremities with a current in the same direction provide a longitudinal confinement. The resulting trap has a cigar shape, with a transverse frequency of a few hundreds of Hz and a longitudinal frequency of a few Hz typically.

Finally, more complex traps have been realised by combining far detuned lasers with electric~\cite{Lemonde1995} or magnetic fields. For example, Bose-Einstein condensation was achieved in the group of Ketterle in 1995 in a plugged quadrupole trap~\cite{Davis1995}, where atoms are prevented from crossing the zero field region by a repulsive blue detuned laser.

\subsection{Evaporative cooling}
\label{sec:evap}
By definition, there is no friction mechanism in a conservative trap. Cooling is nevertheless possible, by limiting the trap depth to a value $U$ adapted to the temperature $T$. As elastic collisions occur, atoms may gain enough energy to escape the trap. The remaining atoms are colder and the phase space density is increased, even if some atoms are lost. This process is called evaporative cooling~\cite{Hess1986,Masuhara1988}. For a fixed depth $U$, the ratio $U/k_B T$ increases as cooling proceeds and the evaporation becomes very slow. To maintain an efficient evaporation, the trap depth has to be decreased with time, keeping the ratio $U/k_B T$ approximately constant~\cite{Luiten1996}.

\begin{figure}
\begin{center}
\begin{tabular}{cl}
\includegraphics[width=40mm]{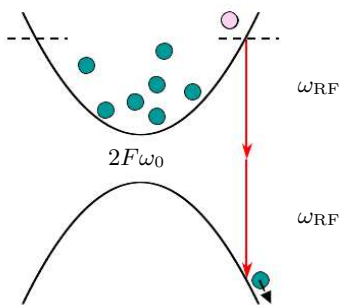} &\\[-35mm]
& $\omega_{\rm RF}$ \\[6mm]
$2F\omega_0$&\\[4mm]
& $\omega_{\rm RF}$ \\[10mm]
&
\end{tabular}
\caption{Principle of evaporative cooling in a magnetic trap. The trap depth in a spin state $m_F$ is limited by the RF field to $m_F \hbar(\omega_{\rm RF} -\omega_0)$ where $\omega_0 = \gamma B_0$ is the resonant frequency at the trap centre.}
\label{fig:evapRF}       
\end{center}
\end{figure}

The implementation of evaporative cooling in a magnetic trap is based on the use of radio-frequency (RF) radiation. The RF induces transitions between magnetic $m_F$ states, such that the atomic spin is flipped to a high-field seeking state, and the atom is expelled from the trap. The transition occurs at the position where the RF is resonant with the magnetic spacing, that is for a given trapping depth, see Fig.~\ref{fig:evapRF}. As temperature decreases, the RF frequency has to be decreased to maintain the cooling efficiency. An RF ramp between a few tens of MHz down to about 1\,MHz is commonly used to reach quantum degeneracy in a magnetically trapped atomic cloud.

Evaporative cooling was also used in dipole traps to reach Bose-Einstein condensation~\cite{Barrett2001}. The laser intensity was decreased with time to reduce the trap depth. Compared to RF evaporation in a magnetic trap, this method has the disadvantage of reducing also the oscillation trapping frequencies which scale as the square root of the laser power, and thus limiting the collision rate. This different scaling~\cite{OHara2001} has to be taken into account in the experiments.

\section{Bose-Einstein condensation}
\subsection{Transition temperature}
Bosons at low temperature undergo a quantum phase transition to a Bose-Einstein condensed state. On the other end, fermions fill the states one by one from the lowest one up to the Fermi level. This strong difference is detectable at low temperature and large density -- that is, at high phase space density $n \lambda_{\rm dB}^3$ -- and is a direct consequence of the different quantum statistics.

The state occupation of bosons follow the Bose-Einstein distribution function:
\begin{equation}
f(\varepsilon) = \frac{1}{e^{(\varepsilon - \mu)/k_B T} - 1}
\end{equation}
where $T$ is the temperature, $\mu$ is the chemical potential and $f(\varepsilon)$ gives the population at energy $\varepsilon$. Above the critical temperature $T_C$, the chemical potential is such that $\mu < E_0$ where $E_0$ is the energy of the ground state, and $f(\varepsilon)$ is always defined even for $\varepsilon = E_0$. Using a semi-classical approximation valid for a large atom number and a level spacing small compared to $k_B T$, the chemical potential is linked to the total atom number $N$ through
\begin{equation}
N= \int_{E_0}^{\infty} \rho(\varepsilon) f(\varepsilon)d\varepsilon
\end{equation}
where $\rho(\varepsilon)$ is the density of states (D.O.S.) at energy $\varepsilon$. For example, in a 3D box of size $L$, $\rho(\varepsilon) = 4\pi (L/h)^3 M^{3/2}\sqrt{2\varepsilon}$. In the following, we will concentrate on the experimentally more relevant case of a 3D harmonic trap of oscillation frequency $\omega_{\rm ho}$, where $\rho(\varepsilon) = \varepsilon^2/\left[2(\hbar\omega_{\rm ho})^3\right]$.

\begin{figure}
\begin{center}
\includegraphics[width=0.8\columnwidth]{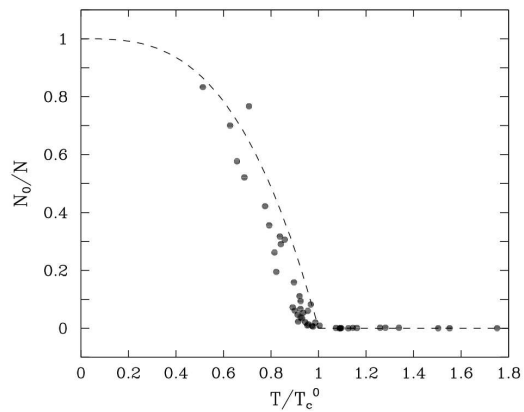}
\caption{Fraction of atoms in the ground state as a function of temperature. Below $T_C$, $N_0/N$ suddenly increases, to reach 1 at $T=0$. Dotted line: prediction of Eq.(\ref{eq:N0surN}). Circles: first experimental data from the JILA group \cite{Ensher1996}. Although the atom number is finite (typically about $10^6$), the transition is quite abrupt and the term `phase transition' is justified. The small shift of the transition temperature is due to atom interactions and finite atom number. Reprinted figure with permission from ref.~\cite{Dalfovo1999}. Copyright (1999) by the American Physical Society. }
\label{fig:N0surN}       
\end{center}
\end{figure}

When $N$ is increased at given $T$, $\mu$ reaches $E_0$ for $N = N_C$ and the ground state population becomes macroscopic. This happens equivalently when $T$ is reduced at given $N$ for a finite critical temperature $T_C$. At this point, the semi-classical approximation remains valid for the excited states only and the ground state population $N_0$ should be counted separately. The total atom number is then $N=N_0 + N'$, where
\begin{equation}
N' = \int_{E_0}^{\infty} \rho(\varepsilon) f(\varepsilon)d\varepsilon \, .
\end{equation}
In a harmonic trap, $N' = \zeta(3) \left[k_B T /(\hbar \omega_{\rm ho})\right]^3$. $\zeta$ is the Riemann zeta function. When this value becomes less than $N$, all the remaining atoms accumulate in the ground state. For $T<T_C$, one thus has:
\begin{eqnarray}
&&\frac{N_0}{N} = 1 - \left(\frac{T}{T_C}\right)^3 \quad , \quad \mbox{with} \label{eq:N0surN}\\
&&k_B T_C = \hbar \omega_{\rm ho} \left( \frac{N}{\zeta(3)} \right)^{1/3} = 0.94 \, \hbar \omega_{\rm ho} N^{1/3} \, .
\label{eq:Tc}
\end{eqnarray}
Below $T_C$, the population of the ground state becomes macroscopic, see Fig.\,\ref{fig:N0surN}. Bose-Einstein condensation appears as a saturation in the excited states population resulting from Bose-Einstein statistics. In particular, the condensation threshold\footnote{Small corrections to Eq.(\ref{eq:Tc}) appear due to finite size effects and atom interactions\,\cite{Dalfovo1999}.} $k_B T_C$ is much larger than the level spacing $ \hbar \omega_{\rm ho}$ for large $N$. Finally, let us point out that in free space --- or in a large box of size $L \rightarrow \infty$ --- the transition temperature is deduced from the famous relation $n \lambda_{\rm dB}^3 = 2.612$.

\subsection{Detection of the BEC}
\begin{figure}
\begin{center}
\includegraphics[width=0.8 \linewidth]{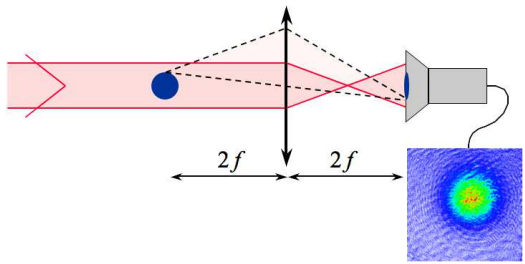}
\caption{Principle of absorption imaging. The shadow of the atomic cloud in a resonant laser beam is imaged with a lens on a CCD camera.}
\label{fig:abs_imaging}       
\end{center}
\end{figure}

In free space -- or in a box -- Bose-Einstein condensation happens in momentum space. In a trap, the condensate wave-function is also localized in space and differs from the non-condensed density. However, a detection of the density profile in the trap is often not used, due to the small size and the very large density ($10^{15}$\,cm$^{-3}$ typically). Rather, an absorption image of the expanding atoms is made after a sudden suppression of the trapping potential. This method is known as time-of-flight imaging. The principle of absorption imaging is depicted on Fig.\,\ref{fig:abs_imaging}.

\begin{figure}[b]
\begin{center}
$(a)$\includegraphics[width=0.3 \columnwidth]{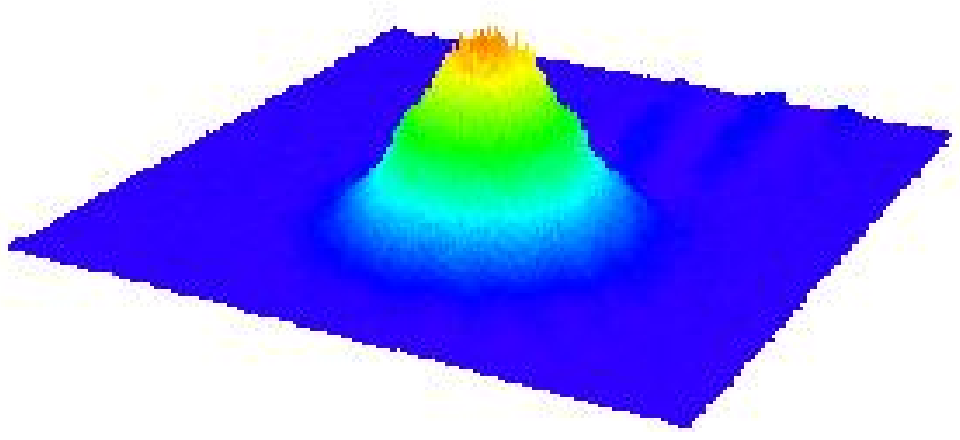}$(b)$\includegraphics[width=0.3 \columnwidth]{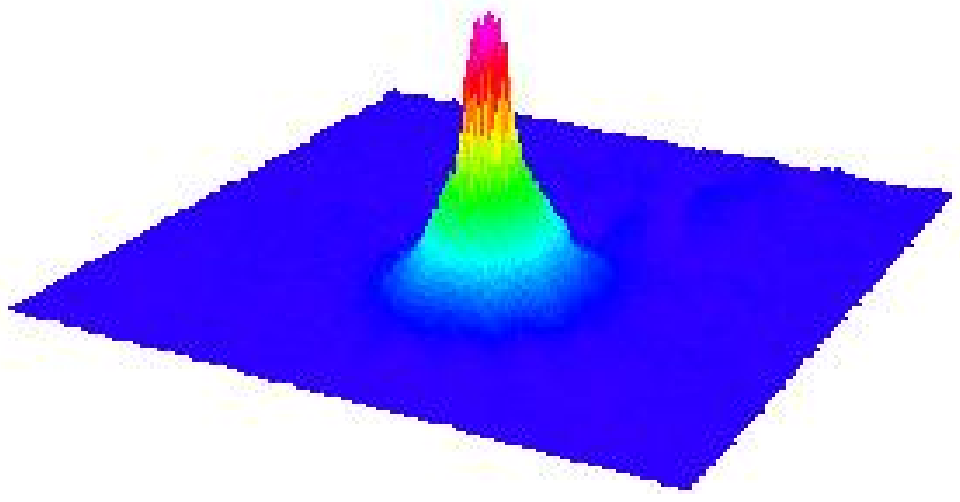}$(c)$\includegraphics[width=0.3 \columnwidth]{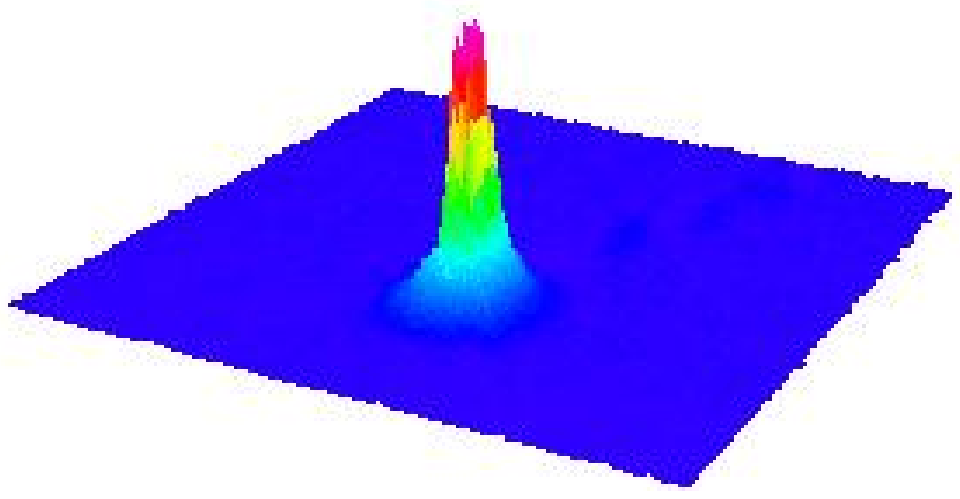}
\caption{Absorption images of an expanding rubidium cloud. $(a)$: thermal gas, $T=1.1~T_C$, $(b)$: double peak structure, with a Bose-Einstein condensate surrounded by a thermal gas, $T=0.6~T_C$ and $(c)$: almost pure condensate, $T=0.3~T_C$. Data taken with rubidium atoms at Paris Nord University.}
\label{fig:onsetBEC}       
\end{center}
\end{figure}
After expansion, the position distribution is essentially a picture of the momentum distribution after all kinetic and interaction energies have been translated into kinetic energy. The thermal cloud has a spherical shape given by the temperature, with a $1/\sqrt{e}$ radius $\Delta p_i = \sqrt{M k_B T}$ in all directions $i=x,y,z$. The condensate part is peaked, with an anisotropic momentum distribution if the trap was anisotropic: $\Delta p_x > \Delta p_y$ if the oscillation frequencies satisfies $\omega_x < \omega_y$. Fig.\,\ref{fig:onsetBEC} presents the density distribution after 20\,ms of expansion at $(a): T=1.1~T_C$, $(b): T=0.6~T_C$ and $(c): T=0.3~T_C$. The distribution is Gaussian above $T_C$, and below $T_C$ the peaked anisotropic condensate density is clearly visible in the middle of a spherical thermal cloud. The two component density distribution is a signature of Bose-Einstein condensation, as observed in the first experiment in JILA~\cite{Anderson1995}.

\subsection{Coherence and atom laser}
Bose-Einstein condensates are often compared to a laser, where the atoms play the role of photons. Indeed, as all atoms populate the same ground state, they are spatially coherent and a constant phase can be attributed to the whole cloud. The coherence length $\lambda_c$, defined as decay length of the first order correlation function, can be measured experimentally in two ways. First, the visibility of matter wave interference fringes in a double slit experiment decreases by a significant amount when the slits are separated by more than $\lambda_c$. The first interference between two independent Bose-Einstein condensates was demonstrated experimentally at MIT in 1997~\cite{Andrews1997}. Later on, quantitative experiments measuring fringe visibility were carried in Munich~\cite{Bloch2000} and at NIST~\cite{Hagley1999}. Second, the momentum distribution is the Fourier transform of the correlation function, and the momentum width is linked to the coherence length through $\Delta p = h/\lambda_c$. Velocity selective Bragg spectroscopy allowed the recording of the momentum distribution~\cite{Stenger1999}. All these experiments concluded that the coherence length of the condensate is equal to its physical size, confirming the coherence properties of the BEC. The coherence length can be limited to a smaller amount for very anisotropic systems, quasi-uni or bidimensional.

\begin{figure}
\begin{center}
\includegraphics[width=0.8 \columnwidth]{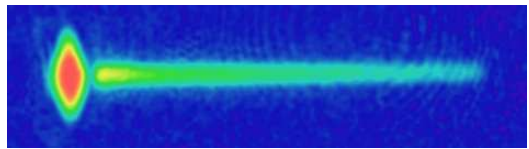}
\caption{A realisation of an atom laser. Rubidium atoms are coupled out of a magnetic trap with a rf field and fall due to gravity (pointing towards the right of the picture). Figure from Bloch, H\"ansch and Esslinger.}
\label{fig:atomlaser}       
\end{center}
\end{figure}

These features make the BEC very promising in atom interferometry, as the laser greatly improves light interferometry. Atom lasers, where a small fraction of the condensed atoms is extracted in a coherent beam with non zero velocity, were already demonstrated~\cite{Bloch1999}, see Fig.\,\ref{fig:atomlaser}. Attempts are made to continuously refill the condensate and realise a continuous atom laser~\cite{Lahaye2004}. For metrological applications to atom interferometry, however, atomic interactions play an important role and should be suppressed or controlled~\cite{Gustavsson2008}.

\subsection{Role of the interactions}
\label{sec:interactions}
What is the many body ground state of the atomic ensemble? In the absence of atomic interaction, we expect a product state of a zero momentum state, or of the harmonic oscillator ground state in a trap. However, even if interactions between atoms have a negligible contribution to the energy above the transition temperature, a variational approach reveals that they have a major influence for the condensate state. Fortunately, the interaction may be described in most cases by a mean field approach, with a single parameter $a$, the scattering length. The condensate wave function $\psi$ obeys the Gross-Pitaevskii equation~\cite{Gross1961,Pitaevskii1961}
\begin{equation}
i \hbar \frac{\partial}{\partial t}\psi(\mathbf{r},t) = \left( -\frac{\hbar^2\nabla^2}{2M} + V(\mathbf{r}) + g|\psi(\mathbf{r},t)|^2 \right) \psi(\mathbf{r},t)
\end{equation}
where $g = 4\pi \hbar^2 a/M$ is the interaction coupling constant. The time-independent Gross-Pitaevskii equation is
\begin{equation}
\left( -\frac{\hbar^2\nabla^2}{2M} + V(\mathbf{r}) + g|\psi(\mathbf{r})|^2 \right) \psi(\mathbf{r}) = \mu \psi(\mathbf{r})
\label{eq:GPE}
\end{equation}
where $\mu$ is the chemical potential.

An analysis of this equation with a Gaussian ansatz for the wave-function shows that to obtain a stable Bose-Einstein condensate, interactions have to be repulsive ($a>0$), such that the attractive trapping force is compensated by a repulsive force between atoms. For $a<0$, the cloud collapses and the condensate is destroyed, unless the atom number is very small and the zero point kinetic energy can compensate both trapping and interactions.

The repulsive interactions are responsible for an increased cloud size with respect to the ground state of the trapping potential. In a harmonic potential of trapping frequency $\omega_{\rm ho}$, the expected size of the ground state in the absence of interaction is $a_{\rm ho} = \sqrt{\hbar/M\omega_{\rm ho}}$ and the corresponding atomic density should be Gaussian. Instead, the interaction term often dominates over the kinetic energy in the Gross-Pitaevskii equation. In this regime, known as the Thomas-Fermi regime and well describing a trapped Bose-Einstein condensate, the laplacian term of Eq.(\ref{eq:GPE}) is dropped and the density is simply an inverted parabola for a harmonic potential:
\begin{equation}
n_0(\mathbf{r}) = |\psi(\mathbf{r})|^2 = \frac{\mu - \frac{1}{2}M\omega_{\rm ho}^2r^2}{g} = \frac{\mu}{g}\left(1 - \frac{r^2}{R^2} \right) \, .
\end{equation}
The Thomas-Fermi radius $R$ and the chemical potential are given by~\cite{Dalfovo1999}
\begin{equation}
\mu = \frac{\hbar\omega_{\rm ho}}{2} \, \left(\frac{15Na}{a_{\rm ho}}\right)^{2/5} \quad \mbox{and} \quad R= a_{\rm ho} \,  \left(\frac{15Na}{a_{\rm ho}}\right)^{1/5} \, .
\end{equation}
$R$ commonly reaches a few times $a_{\rm ho}$
\begin{figure}
\begin{center}
\includegraphics[width=0.8 \columnwidth]{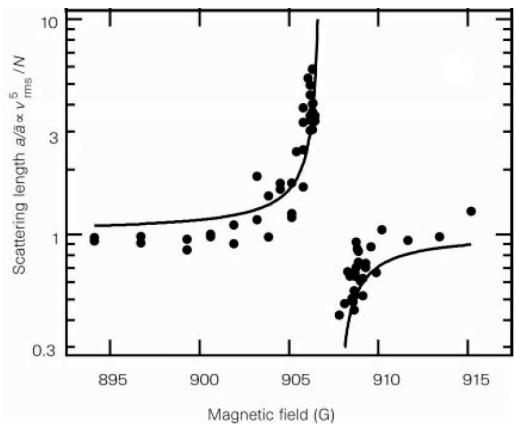}
\caption{Feshbach resonance in sodium. The scattering length diverges and changes its sign around a resonant magnetic files. Adapted by permission from Macmillan Publishers Ltd.: Nature \cite{Inouye1998}, Copyright (1998).}
\label{fig:Feshbach}       
\end{center}
\end{figure}

A very attractive characteristic of quantum degenerate gases is their flexibility. For example, atoms can be trapped in lattices of tunable depth and period, the temperature can be adjusted, etc. The strength of the interactions is also adjustable in many cases by the use of a Feshbach resonance~\cite{Inouye1998} as shown in Fig.\,\ref{fig:Feshbach}. The scattering length $a$ then depends on the magnetic field, diverging and changing its sign for a given value of the magnetic field. This allows a fine tuning of this important parameter. The control over the interactions was an important element in the successful observation of Bose condensed caesium~\cite{Weber2003}, an atom for which the collisional properties are not favourable. In chromium, the suppression of contact interactions allowed the enhancement of dipolar interactions between atoms~\cite{Lahaye2007}, which opens the way to the realisation of magnetic model systems for condensed matter. Finally, scanning the magnetic field through a Feshbach resonance results in atoms pairing into diatomic molecules. In a mixture of degenerate gases, heteronuclear molecules can be produced efficiently~\cite{Ospelkaus2006}.

\subsection{BEC as a model system}
Bose-Einstein condensates are highly controllable. As described in section~\ref{sec:interactions}, the interparticle interactions can be tuned in many cases. Moreover, weakly interacting degenerate gases are model systems for condensed matter physics in many respects.

\subsubsection{Superfluidity}
\begin{figure}
\begin{center}
\includegraphics[width=0.8 \columnwidth]{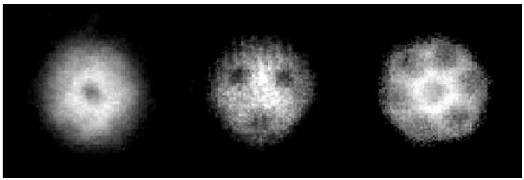}
\caption{Evidence of vortex formation in a rotating Bose-Einstein condensate. The vortices are produced by stirring the atoms with a rotating laser beam. From left to right: 1, 3 and 7 vortices are detected after a time-of-flight expansion. The number of vortices increases with the stirring frequency. Figure by Chevy, Madison, Wohlleben and Dalibard.}
\label{fig:vortices}       
\end{center}
\end{figure}

Bose-Einstein condensates with a finite scattering length $a$ are superfluids. Indeed, the Gross-Pitaevskii equation was introduced for superfluid helium~\cite{Gross1961,Pitaevskii1961} where interactions are strong, but is much more accurate for Bose condensates. An evidence for superfluidity is the observation for vortices in a rotating BEC~\cite{Matthews1999,Madison2000}, where the orbital angular momentum is quantized in units of $\hbar$ around a vortex line, see Fig.\,\ref{fig:vortices}. Recently, a persistent atomic current was observed in a circular potential~\cite{Ryu2007}. This characteristic makes the BEC a model system for superfluidity.

\begin{figure}
\begin{center}
\includegraphics[width=0.8 \columnwidth]{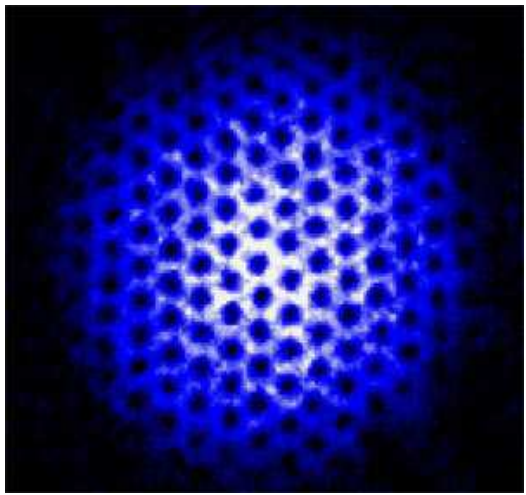}
\caption{Absorption images of an expanding vortex triangular lattice, or Abrikosov lattice, taken in the group of Cornell (JILA, Boulder). The atomic density drops in each vortex core.}
\label{fig:Abrikosov}       
\end{center}
\end{figure}

A clear signature of superfluidity in Bose-Einstein condensates is the observation of vortices in a rotating degenerate gas. Indeed, the Gross-Pitaevskii equation is equivalent to hydrodynamical equations for the density $n$ and the velocity field $\mathbf{v}$:
\begin{eqnarray}
&& \frac{\partial n}{\partial t} + \nabla \cdot (n\mathbf{v}) = 0 \, ,\label{eq:mass_conservation}\\
&&M\left(  \frac{\partial \mathbf{v}}{\partial t} + \frac{1}{2} \nabla v^2 \right) = \nabla \left( \frac{\hbar^2}{2M} \frac{{\rm \Delta} (\sqrt{n})}{\sqrt{n}} - V(\mathbf{r}) - gn \right) \, , \label{eq:Euler}
\end{eqnarray}
where the velocity field is defined by
\begin{equation}
\psi(\mathbf{r},t) = \sqrt{n(\mathbf{r},t)} \ e^{i\phi(\mathbf{r},t)} \quad \mbox{and} \quad \mathbf{v}(\mathbf{r},t) = \frac{\hbar}{M} \nabla \phi(\mathbf{r},t) \, .
\end{equation}
Eq.(\ref{eq:mass_conservation}) and (\ref{eq:Euler}) account for mass conservation and the Euler equation, respectively. The velocity is directly related to the phase gradient. As a consequence, the flow is irrotational. The circulation of the velocity field is an integer multiple of $h/M$. When the condensate is put into rotation, vortex lines appear above a critical rotation frequency and organise themselves into an Abrikosov lattice~\cite{Madison2000,AboShaeer2001}. The velocity decreases as $1/r$ from the vortex centre, whereas the density drops inside the vortex core, as can clearly be seen on absorption pictures, Fig.~\ref{fig:Abrikosov}.

Quantum degenerate gases are also a model for superconductivity. Recently, BCS pairing was achieved in quantum degenerate fermions by the use of Feshbach resonances~\cite{Chin2004}. On one side of the resonance, where the interactions are repulsive, the atoms are paired into bosonic molecules and form a BEC, whereas they form Cooper pairs when the interactions are attractive. Both molecular BECs and BCS fluids are superfluids and vortices are formed by rotating the sample~\cite{Zwierlein2005}.

\subsubsection{Josephson oscillations}
A double well potential was realised with dipole traps by Oberthaler and co-workers~\cite{Albiez2005} to control the tunnel coupling $J$ between two small atom traps. This system is a model of Josephson junctions in solid state physics (see the review~\cite{QMJeanneret} and references therein), where the atom difference -- and thus the chemical potential difference -- plays the role of the voltage applied to the junction. Two traps containing $N_a$ and $N_b$ atoms respectively can exchange particles by tunneling through a barrier. The relative phase between the two wells $\phi$ and the atom number difference $n=N_a - N_b$ are conjugate variables, and obey the Josephson equations
\begin{equation}
\hbar \frac{\partial n}{\partial t} = J \sin\phi \quad , \quad \frac{\partial \phi}{\partial t} = -\frac{U}{\hbar}n \, .
\end{equation}
These two equations are similar to those of a classical pendulum. At small difference number $n$, spontaneous oscillation of the atom number occur between the two wells. For a marked initial asymmetry, the chemical potentials differ by more than the tunnel coupling and $n$ is stationary: the situation is known as self-trapping.

Interactions can be used in this system to limit the relative atom number fluctuations between the wells to better than $1/\sqrt{N}$, the limit given by Poissonian noise. As phase difference and number difference are conjugate variables, the Heisenberg limit is expected to be of order $1/N$. Such a reduction of the fluctuations in the difference of atom number is called squeezing. Recent experiments in the same group have demonstrated number squeezing by measuring independently the relative phase fluctuations and the number fluctuations~\cite{Esteve2007}.

\subsubsection{Optical lattices}
\begin{figure}
\begin{center}
\includegraphics[width=0.8 \columnwidth]{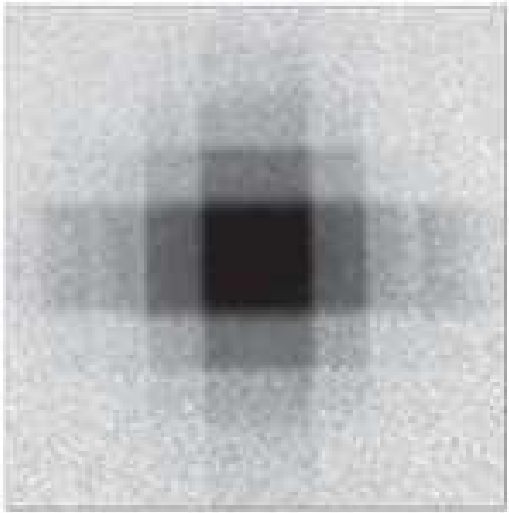}
\caption{Population in the Brillouin zones for a cubic lattice, imaged by the technique of adiabatic mapping, \textit{i.e.} adiabatic switching of the optical lattice. Adapted by permission from Macmillan Publishers Ltd.: Nature Physics \cite{Bloch2005}, Copyright (2005).}
\label{fig:Brillouin}       
\end{center}
\end{figure}

To make ultracold atoms even more similar to electrons in solids, optical lattices are widely used in BEC experiments, see section~\ref{sec:dipoleforce}. Bloch oscillations in optical lattices were observed already in 1995 with Raman cooled atoms~\cite{Dahan1995}, and are used in metrology experiments for the determination of the fine structure constant~\cite{QMGuellati}. In the lattice, atoms occupy Bloch states in the lower band. The wave function is delocalised over many lattice sites and the momentum distribution is a superposition of peaks spaced by $\hbar/d$ where $d$ is the lattice period. Time of flight experiments, when the lattice is switched off abruptly, show these periodic structures in momentum, resulting from the interference between the different lattice sites. If on the other hand the lattice is ramped down adiabatically, the Brillouin zones can be imaged in a time of flight experiment, see Fig.\,\ref{fig:Brillouin}, allowing the measurement of the atomic population in each Bloch band~\cite{Bloch2005}.

In this system, the interplay between tunneling and interactions provides rich physical phenomena.
When interactions become important, the condensed atoms can no longer be described as a simple matter wave, and become strongly correlated. The superfluid to Mott insulator transition is an example of this transition from weakly interacting to strongly correlated system. It was demonstrated in 2002 with rubidium atoms in an optical lattice, in the group of H\"ansch~\cite{Greiner2002}. Again, this quantum phase transition was described theoretically in the framework of condensed matter systems like liquid $^4$He or superconducting materials~\cite{Fischer1989}.

The system is well described by the Bose-Hubbard hamiltonian:
\begin{equation}
\hat{H}=-J\sum_{<i,j>}\hat{a}^+_i \hat{a}_j + \frac{U}{2} \sum_i \hat{n}_i (\hat{n}_i -1) \, .
\end{equation}
The first term describes tunneling from site $j$ to site $i$ with amplitude $J$, the sum being performed only over neighbouring sites $<i,j>$. The second term accounts for on-site interactions. For contact interactions, interaction between atoms further apart than one site is negligible.

Depending on the ratio of the tunneling energy to the interaction energy $J/U$, the system changes from phase coherent to strongly correlated. The transition is driven by varying the standing wave intensity and thus the lattice depth. For a small lattice depth, the on-site interaction energy $U$ is low and the tunneling is large, and tunneling ensures phase coherence throughout the sample. Time of flight experiments reveal the usual interference pattern between the sites. In this regime, the number of atoms per lattice site fluctuates. For deeper optical lattices, the energy cost of having a different population in different lattice sites becomes too large, and the atom number per site is locked to an integer value. Phase coherence is then lost and the interference pattern disappears. Phase difference and atom number difference are conjugated variables in this problem, as they are in the case of the double well. The Mott insulating state was also observed very recently in a degenerate Fermi gas~\cite{Joerdens2008}.

The Bose-Hubbard hamilonian describes a system with contact interactions. In 2005, a Bose-Einstein condensate of chromium was obtained in Stuttgart~\cite{Griesmaier2005}, and very recently at Paris Nord~\cite{Beaufils2008}. Chromium is of particular interest because its magnetic moment is 6 times larger than for alkali atoms. As a consequence, dipolar interactions are 36 times larger, and their effect have been evidenced in several experiments in Stuttgart. Together with the control of contact interactions, which can be minimized to enhance the effect of dipolar interactions, this opens the ways to manipulate quantum ferrofluids~\cite{Lahaye2007}. When stored in an optical lattice, quantum phase transitions could be observed between different magnetic states.

\subsubsection{And even more...}
Bose-Einstein condensates have already proven to be a source of inspiration for theoreticians with a background in condensed matter, and reciprocally condensed matter has inspired many beautiful experiments with ultracold gases. I will not give an extensive list of example of such fruitful interactions between the two fields. Let me however cite a few more situations where these interactions are at work.

In real condensed matter systems, disorder is always present to some extent. Now, disorder can also be mimicked for Bose-Einstein condensates, by using pseudo-periodic potentials in a lattice geometry~\cite{Fallani2007} or superimposing a light speckle pattern to a trapped sample. A transition from a superfluid to an insulating state is also induced by disorder. This was demonstrated very recently in an experiment where disorder was introduced in one dimension, and Anderson localisation of matter waves was demonstrated~\cite{Billy2008}.

Standing waves are used to produce optical lattices. When the lattice is realised only in one or two dimensions, it results in a collection of anisotropic traps, where the atoms live in 2D or 1D respectively. New effects appear in low dimensional systems~\cite{Pricoupenko2004}. In two dimensions, Bose-Einstein condensation is possible only in a trap. When the temperature is raised from zero, the gas then enters the Berezinskii-Kosterlitz-Thouless phase (BKT), where vortex-antivortex pairs are present~\cite{BKT}. These pairs are destroyed as the temperature increases further, and superfluidity is lost. The BKT transition was recently observed in a 2D rubidium gas~\cite{Hadzibabic2006}. In one dimension, the system becomes strongly interacting when the density is reduced. The gas then enters the Tonks-Girardeau regime, where correlations imply a fermionization of the system~\cite{GIrardeau1960}: the strong interactions act as exchange interaction and prevent two atoms from sitting at the same position. Tonks-Girardeau gases were also observed with ultracold bosons recently~\cite{Paredes2004,Kinoshita2004}.

Finally, fermionic degenerate gases are also widely investigated. In these systems, BCS-like pairing was observed~\cite{Chin2004,Zwierlein2005}. Pairing in a situation were the spin up and spin down populations are not balanced is also studied. There is a hope that these experiments will give key explanations of high $T_C$ superconductivity. More generally, applications of ultracold atoms in atom interferometry, simulation of quantum matter or quantum computation have just started and are progressing rapidly.


\begin{thebibliography}{99}
\bibitem{Hansch1975} T.W. H{\"ansch} and A.L. Schawlow, Opt. Comm. \textbf{13}, (1975) 393
\bibitem{QMBize} S. Bize et al., At. Mol. Opt. Phys. \textbf{42}, 95 (2000)
\bibitem{QMLemonde} P. Lemonde, Eur. Phys. J. Special Topics \textbf{172}, 81 (2009)
\bibitem{QMMargolis} H. Margolis, Eur. Phys. J. Special Topics \textbf{172}, 97 (2009)
\bibitem{QMGuellati} M. Cadoret, E. De Mirandes, P. Clad\'e, F. Nez, L. Julien, F. Biraben, and S. Guellati-Khelifa, Eur. Phys. J. Special Topics \textbf{172}, 121 (2009)
\bibitem{Anderson1995} M.\,H.~Anderson, J. R. Ensher, M.R. Matthews, C.E. Wieman, and E.A. Cornell, Science \textbf{269}, (1995) 198
\bibitem{Cornell2002} E.~A.~Cornell and C.~E.~Wieman, Rev. Mod. Phys. \textbf{74}, (2002) 875
\bibitem{Ketterle2002} W.~Ketterle, Rev. Mod. Phys. \textbf{74}, (2002) 1131
\bibitem{DeMarco1999} B. DeMarco and D.S. Jin, Science \textbf{285}, (1999) 1703
\bibitem{QMJeanneret} B. Jeanneret and S.P. Benz, Eur. Phys. J. Special Topics \textbf{172}, 181 (2009)
\bibitem{QMKemppinen} A. Kemppinen, M. Meschke, M. M\"ott\"onen, D. V. Averin, and J. P. Pekola Eur. Phys. J. Special Topics \textbf{172}, 311 (2009)
\bibitem{QMGallop} J. Gallop, Eur. Phys. J. Special Topics \textbf{172}, 399 (2009)
\bibitem{Chabe2007} J. Chab\'e, G. Lemari\'e, B. Gr\'emaud, D. Delande, P. Szriftgiser, and J.-C. Garreau, Preprint arXiv:0709.4320 (2007) 
\bibitem{Billy2008}
J. Billy, V. Josse, Z. Zuo, A. Bernard, B. Hambrecht, P. Lugan, D. Cl\'ement, L. Sanchez-Palencia, P. Bouyer, and A. Aspect, Nature \textbf{453}, 891 (2008)
\bibitem{LesHouches90}
\textit{Fundamental systems in quantum optics}, proceedings of the Les Houches LIII Summer School, edited by J. Dalibard, J.-M. Raymond and J. Zinn-Justin (North Holland, 1992); see in particular the courses by C. Cohen-Tannoudji, p. 1, and W. D. Phillips, p. 165
\bibitem{Varenna98}
\textit{Bose-Einstein Condensation in Atomic Gases}, Proceedings of the International School of Physics (Enrico Fermi), Course CXl, edited by M. Inguscio, S. Stringari and C. E. Wieman (IOS Press, Amsterdam, 1999)
\bibitem{Dalfovo1999}
F. Dalfovo, S. Giorgini, L. P. Pitaevskii, and S. Stringari,
Rev. Mod. Phys. \textbf{71}, (1999) 463
\bibitem{LesHouches99}
\textit{Coherent matter waves}, proceedings of the Les Houches LXXII Summer School, edited by R. Kaiser, C. Westbrook and F. David (Springer, 2001)
\bibitem{Pethick2002} C. J. Pethick and H. Smith, \textit{Bose-Einstein Condensation in Dilute Gases} (Cambridge University Press, Cambridge 2002)
\bibitem{Pitaevskii2003} L. P. Pitaevskii and S. Stringari, \textit{Bose-Einstein Condensation} (Oxford University Press, Oxford 2003)
\bibitem{CohenCollege} C. Cohen-Tannoudji, lectures at Coll\`ege de France (in french)
\bibitem{Metcalf} H. J. Metcalf and P. van der Straten, \textit{Laser cooling and Trapping} (Springer, New York 1999)
\bibitem{CohenNobel} C. Cohen-Tannoudji, Rev. Mod. Phys. \textbf{70}, (1998) 707
\bibitem{ChuNobel} S. Chu, Rev. Mod. Phys. \textbf{70}, (1998) 685
\bibitem{PhillipsNobel} W. D. Phillips, Rev. Mod. Phys. \textbf{70}, (1998) 721
\bibitem{Prodan1985} J. Prodan, A. Migdall, W. D. Phillips, I. So, H. Metcalf and J. Dalibard, Phys. Rev. Lett. \textbf{54}, (1985) 992
\bibitem{Chu1986} S. Chu, J. E. Bjorkholm, A. Ashkin, and A. Cable, Phys. Rev. Lett. \textbf{57}, (1986) 314
\bibitem{Cook1982} R.J. Cook and R.K. Hill, Opt. Commun. \textbf{43}, (1982) 258
\bibitem{Balykin1987} V.I. Balykin, V.S. Letokhov, Yu.B. Ovchinnikov, and A.I. Sidorov, JETP Lett. \textbf{45}, (1987) 353; Phys. Rev. Lett. \textbf{60}, (1988) 2137
\bibitem{Grimm2000} R. Grimm, M. Weidem\"uller, and Yu.B. Ovchinnikov, Adv. At. Mol. Opt. Phys. \textbf {42}, (2000) 95
\bibitem{Westbrook1990} C.I. Westbrook, R.N. Watts, C.E. Tanner, S.L. Rolston, W.D. Phillips, P.D. Lett and P.L. Gould, Phys. Rev. Lett. \textbf{65}, (1990) 33
\bibitem{Grynberg2001} G. Grynberg and C. Robilliard,
Phys. Rep. \textbf{355}, (2001) 335
\bibitem{Bloch2005} For a review, see I. Bloch, Nature Physics \textbf{1}, (2005) 23
\bibitem{Chu1985} S. Chu, L. Hollberg, J. E. Bjorkholm, A. Cable, and A. Ashkin, Phys. Rev. Lett. \textbf{55}, (1985) 48
\bibitem{Letokhov1977} V.S. Letokhov, V.G. Minogin, and B.D. Pavlik, Sov. Phys. JETP Lett. \textbf{45}, (1977) 698; D. Wineland and W. Itano, Phys. Rev. A \textbf{20}, (1979) 1521
\bibitem{Lett1988} P. D. Lett, R. N. Watts, C. I. Westbrook, W. D. Phillips, P. L. Gould and H. J. Metcalf, Phys. Rev. Lett. \textbf{61}, (1988) 169
\bibitem{Dalibard1989} J. Dalibard and C. Cohen-Tannoudji,
J. Opt. Soc. Am. B \textbf{6}, (1989) 2023; P. J. Ungar, D. S. Weiss, E. Riis, and S. Chu,
\textit{ibid.} 2058
\bibitem{Aspect1988} A. Aspect, E. Arimondo, R. Kaiser, N. Vansteenkiste and C. Cohen-Tannoudji, Phys. Rev. Lett. \textbf{61}, (1988) 826
\bibitem{Kasevich1992} M. Kasevich and S. Chu, Phys. Rev. Lett. \textbf{69}, (1992) 1741
\bibitem{Wineland1975} D. Wineland and H. Dehmelt, Bull. Am. Phys. Soc. \textbf{20}, (1975) 637
\bibitem{Diedrich1989} F. Diedrich, J. C. Bergquist, W. M. Itano, and D. J. Wineland, Phys. Rev. Lett. \textbf{62}, (1989) 403
\bibitem{Raab1987} E. Raab, M. Prentiss, A. Cable, S. Chu, and D. Pritchard,
Phys. Rev. Lett. \textbf{59}, (1987) 2631
\bibitem{Kerman2000} A. J. Kerman, V. Vuletic, C. Chin, and S. Chu, Phys. Rev. Lett. \textbf{84}, (2000) 439
\bibitem{Han2000} D. J. Han, S. Wolf, S. J. Oliver, C. McCormick, M. T. DePue and D. S. Weiss,
Phys. Rev. Lett. \textbf{85}, (2000) 724
\bibitem{Barrett2001} M. D. Barrett, J. A. Sauer, and M. S. Chapman,
Phys. Rev. Lett. \textbf{87}, (2001) 010404
\bibitem{Pritchard1983} D. E. Pritchard, Phys. Rev. Lett. \textbf{51}, (1983) 1336
\bibitem{Lemonde1995}  P. Lemonde, O. Morice, E. Peik, J. Reichel, H. Perrin, W. H\"ansel and C. Salomon, Europhys. Lett. \textbf{32}, (1995) 555
\bibitem{Davis1995} K. B. Davis, M.-O. Mewes, M. R. Andrews, N. J. van Druten, 
D. S. Durfee, D. M. Kurn, and W. Ketterle,
Phys. Rev. Lett. \textbf{75}, (1995) 3969
\bibitem{Hess1986} H. F. Hess, Phys. Rev. B \textbf{34}, (1986) 3476
\bibitem{Masuhara1988} N. Masuhara, J. M. Doyle, J. C. Sandberg, D. Kleppner, T.J. Greytak, H. F. Hess, and G. P. Kochanski, Phys. Rev. Lett. \textbf{61}, (1988) 935
\bibitem{Luiten1996} O. J. Luiten, M. W. Reynolds, and J.T.M. Walraven, Phys. Rev. A \textbf{53}, (1996) 381
\bibitem{OHara2001} K. M. O'Hara, M. E. Gehm, S. R. Granade, and J. E. Thomas, Phys. Rev. A \textbf{64}, (2001) 051403
\bibitem{Ensher1996} J.R. Ensher, D.S. Jin, M.R. Matthews, C.E. Wieman, and E.A. Cornell, Phys. Rev. Lett. \textbf{77}, (1996) 4984
\bibitem{Andrews1997} M. R. Andrews,    C. G. Townsend,    H.-J. Miesner,    D. S. Durfee,    D. M. Kurn,  and  W. Ketterle,
Science \textbf{275}, (1997) 637
\bibitem{Bloch2000} I. Bloch, T.W. H\"ansch and T. Esslinger,
Nature \textbf{403}, (2000) 166
\bibitem{Hagley1999} E.W. Hagley, L. Deng, M. Kozuma, M. Trippenbach, Y. B. Band, M. Edwards, M. Doery, P. S. Julienne, K. Helmerson, S. L. Rolston, and W. D. Phillips,
Phys. Rev. Lett. \textbf{83}, (1999) 3112
\bibitem{Stenger1999} J. Stenger, S. Inouye, A. P. Chikkatur, D. M. Stamper-Kurn, D. E. Pritchard, and W. Ketterle,
Phys. Rev. Lett. \textbf{82}, (1999) 4569
\bibitem{Bloch1999} I. Bloch, T.W. H\"ansch and T. Esslinger,
Phys. Rev. Lett. \textbf{82}, (1999) 3008
\bibitem{Lahaye2004} T. Lahaye, J. M. Vogels, K. Guenter, Z. Wang, J. Dalibard, and D. Gu\'ery-Odelin,
Phys. Rev. Lett. \textbf{93}, (2004) 093003
\bibitem{Gustavsson2008} M. Gustavsson, E. Haller, M. J. Mark, J. G. Danzl, G. Rojas-Kopeinig, and H.-C. N\"agerl,
Phys. Rev. Lett. \textbf{100}, (2008) 080404
\bibitem{Gross1961} E.P. Gross, Nuovo Cimento \textbf{20}, (1961)  454; J. Math. Phys. \textbf{4}, (1963) 195
\bibitem{Pitaevskii1961} L. P. Pitaevskii, Zh. Eksp. Teor. Fiz. \textbf{40}, (1961) 646 [Sov. Phys. JETP \textbf{13}, (1961) 451]
\bibitem{Inouye1998} S. Inouye, M.R. Andrews, J. Stenger, H.-J. Miesner, D.M. Stamper-Kurn, and W. Ketterle,
Nature \textbf{392}, (1998) 151
\bibitem{Weber2003} T. Weber, J. Herbig, M. Mark, H.-C. N\"agerl, and R. Grimm, Science \textbf{299}, (2003) 232
\bibitem{Lahaye2007} T. Lahaye, T. Koch, B. Fr\"ohlich, M. Fattori, J. Metz, A. Griesmaier, S. Giovanazzi, T. Pfau,
Nature \textbf{448}, (2007) 672
\bibitem{Ospelkaus2006}
C. Ospelkaus, S. Ospelkaus, L. Humbert, P. Ernst, K. Sengstock, and K. Bongs, Phys. Rev. Lett. \textbf{97}, (2006) 120402
\bibitem{Matthews1999}
M. R. Matthews, B. P. Anderson *, P. C. Haljan, D. S. Hall †, C. E. Wieman, and E. A. Cornell, Phys. Rev. Lett. \textbf{83}, (1999) 2498
\bibitem{Madison2000}
K. W. Madison, F. Chevy, W. Wohlleben, and J. Dalibard, Phys. Rev. Lett. \textbf{84}, (2000) 806
\bibitem{Ryu2007}
C. Ryu, M. F. Andersen, P. Clad\'e, V. Natarajan, K. Helmerson, and W. D. Phillips, Phys. Rev. Lett. \textbf{99}, 260401 (2007)
\bibitem{AboShaeer2001} J. R. Abo-Shaeer, C. Raman, J. M. Vogels, and W. Ketterle,
Science \textbf{292}, (2001) 476
\bibitem{Chin2004}
C. Chin, M. Bartenstein, A. Altmeyer, S. Riedl, S. Jochim, J. Hecker Denschlag, and R. Grimm,
Science \textbf{305}, (2004) 1128
\bibitem{Zwierlein2005}
M.W. Zwierlein, J.R. Abo-Shaeer, A. Schirotzek, C.H. Schunck, and W. Ketterle,
Nature \textbf{435},  (2005) 1047
\bibitem{Albiez2005} M. Albiez, R. Gati, J. Fölling, S. Hunsmann, M. Cristiani and M.K. Oberthaler,
Phys. Rev. Lett. \textbf{95}, (2005) 010402
\bibitem{Esteve2007} J. Est\`eve and M.K. Oberthaler, private communication
\bibitem{Dahan1995} M. Ben Dahan, E. Peik, J. Reichel, Y. Castin, and C. Salomon, Phys. Rev. Lett. \textbf{76}, (1996) 4508
\bibitem{Greiner2002} M. Greiner, O. Mandel, T. Esslinger, T.W. H\"ansch, and I. Bloch,
Nature \textbf{415}, (2002) 39
\bibitem{Fischer1989} M. P. A. Fischer, P. B. Weichman, G. Grinstein, and D. S. Fischer,
Phys. Rev. B \textbf{40}, (1989) 546
\bibitem{Joerdens2008} R. J\"ordens, N. Strohmaier, K. G\"unther, H. Moritz, and T. Esslinger,
Preprint arXiv:0804.4009 (2008)
\bibitem{Griesmaier2005} A. Griesmaier, J. Werner, S. Hensler, J. Stuhler, and T Pfau,
Phys. Rev. Lett. \textbf{94}, (2005) 160401
\bibitem{Beaufils2008} Q. Beaufils, R. Chicireanu, T. Zanon, B. Laburthe-Tolra, E. Mar\'echal, L. Vernac, J.-C. Keller, and O. Gorceix,
to appear in Phys. Rev. A, Preprint arXiv:0712.3521 (2007)
\bibitem{Fallani2007} L. Fallani, J. E. Lye, V. Guarrera, C. Fort, and M. Inguscio, Phys. Rev. Lett. \textbf{98}, (2007) 130404
\bibitem{Pricoupenko2004} \textit{Quantum Gases in Low Dimensions}, proceedings of the Les Houches school QGLD 2003, edited by L. Pricoupenko, H. Perrin and M. Olshanii, J. Phys. IV \textbf{116} (EDP Sciences, 2004)
\bibitem{BKT} V. L. Berezinskii,
Sov. Phys. JETP \textbf{34}, (1971) 610; J. M. Kosterlitz and D. J. Thouless,
J. Phys. C: Solid State Phys. \textbf{6}, (1973) 1181
\bibitem{Hadzibabic2006} Z. Hadzibabic, P. Kr\"uger, M. Cheneau, B. Battelier, and J. Dalibard,
Nature \textbf{441}, (2006) 1118
\bibitem{GIrardeau1960} M. Girardeau,
J. Math. Phys. \textbf{1}, (1960) 516 
\bibitem{Paredes2004} B. Paredes, A. Widera, V. Murg, O. Mandel, S. F\"olling, I. Cirac, G. V. Shlyapnikov, T. W. H\"ansch, and I. Bloch,
Nature \textbf{429}, (2004) 277
\bibitem{Kinoshita2004} T. Kinoshita, T. Wenger, and D. S. Weiss,
Science \textbf{305}, (2004) 1125
\end{thebibliography}
\end{document}